# Focus optimization in a Computational Confocal Microscope

Keith Dillon

We consider the numerical optimization of performance for a computational extension of a confocal microscope. Using a system where the pinhole detector is replaced with a detector array, we seek to exploit this additional information for each point in the scan. We derive an optimal estimate of the light at focus which minimizes the contribution of out-of-focus light. We estimate the amount of improvement that would be theoretically possible in point-scanning and line-scanning systems and demonstrate with simulation. We find that even with a large degree of regularization, a significant improvement is possible, especially for line-scanning systems.

## Introduction

Confocal microscopes perform optical depth sectioning [1] by scanning a beam focused at the desired depth with a high numerical aperture objective. The transmitted or scattered (depending on mode of operation) light is focused onto a pinhole aperture to reject return from out-of-focus points on the sample, before being detected, one pixel per step in the scan. The effect is that the scattering and absorption properties for each pixel in the final image are limited to a small spatial region defined by the product of the transfer functions of the initial and final objectives. Computational confocal microscopy [2] is a variation on this system which replaces the pinhole with a detector array. If a phase detection method is employed, then this detection may be performed equivalently in the image or Fourier domain. This enables simultaneous depth sectioning of not only scattering but also attenuation and optical path delay [3,4]. Successful application of this approach to confocal endoscopy is demonstrated in [5].

In this paper we consider the improvement that may be gained in such a computational imaging system by optimizing use of the detector samples. The goal is reducing the image degradation due to out-of-focus scattered light. We cast the problem of improving the image as an optimization of the point-spread function (PSF) for a single point in the confocal scan. We view the estimate as a computational point-spread function.

A related category is deconvolution microscopy [6], wherein confocal microscopy is addressed in a subcategory of methods. Deconvolution microscopy techniques can be divided into those that assume knowledge of the 3D point-spread function (the assumption made here), and those that do not, called blind deconvolution. Applied to confocal imaging, deconvolution microscopy uses the pinhole measurements for the entire scan, which may be over the focal plane or over a series of different focal planes. This paper will instead assume that points neighboring the pinhole are also collected for processing. This ceases to be a simple deconvolution problem as each different detector point has a different PSF. For the point in the detector array which might be considered the original pinhole, the focal points for the illumination and detection are aligned, but for other detector points the detection focus is shifted. And the PSF is essentially the product of these two transfer functions.

More closely related techniques are those confocal systems where more detection information is used beyond a single pinhole. In differential detection we have the subtraction of a pair of neighboring detector points, which would be expected to reduce the out-of-focus light more than the focus light if the separation between the detector points are chosen carefully. In [7] a pair of neighboring detection lines was used in a slit-scanning confocal system. The researchers subtracted one half the signal at the offset lines form the central line to achieve a significant reduction in out-of-focus light. Successful implementation of this approach is demonstrated in [8,9].

Such methods demonstrate obvious qualitative improvements in image contrast, however the image itself is now a high-pass-filtered version of the truth. Measurements of out-of-focus light which use the integration over planes at different defocus will ignore this fact, and further will incorrectly attribute light at the focal plane from adjacent points in the transverse direction, as light in focus, potentially causing this metric to overestimate the improvement. One might assume that this issue may be addressed by a deconvolution of the subsequent high-pass-filtered image to remove the filtering effect in order to make a fair comparison, though the final performance remains to be demonstrated. The estimate derived in this paper may be considered a way to simultaneously minimize the effect of light from points off the focal plane axially, as well as transverse to the focus. We also demonstrate numerically the performance of this optimum.

## Method

We use Debye theory [10] to numerically estimate the field of a high numerical-aperture objective using a collection of plane waves over the solid angle $\Omega$ produced by the aperture, yielding

$$U(\mathbf{r}) = \frac{i}{\lambda} \iint_\Omega \sqrt{\cos\theta}\, e^{-k\mathbf{s}\cdot\mathbf{r}} d\Omega. \tag{1}$$

We used apodization based on the sine condition, and an otherwise constant aperture function. For a confocal system, we use this field estimate as the complex transfer function for both objectives,

$$T_i(\mathbf{r}) = T_d(\mathbf{r}) = U(\mathbf{r}). \tag{2}$$

$T_i$ and $T_d$ are the illumination and detection transfer functions respectively. The transfer function of the confocal system is the product of the illumination and detection transfer functions. And the (infinitesimal) pinhole signal is, assuming incoherence between light from different sample points,

$$\begin{aligned}I_0 &= \iiint_\mathbf{r} |T_i(\mathbf{r}) T_d(\mathbf{r}) f(\mathbf{r})|^2 d\mathbf{r} \\ &= \iiint_\mathbf{r} OTF(\mathbf{r}) |f(\mathbf{r})|^2 d\mathbf{r}\end{aligned} \tag{3}$$

where $f(\mathbf{r})$ is the object and $OTF(\mathbf{r})$ is the optical transfer function of the microscope. We describe the scanning of the sample as the translation of both illumination and detection transfer functions relative to the sample,

$$I_0 = \iiint_{\mathbf{r}} |T_i(\mathbf{r}+\boldsymbol{\Delta})T_d(\mathbf{r}+\boldsymbol{\Delta})f(\mathbf{r})|^2 d\mathbf{r}$$
$$= \iiint_{\mathbf{r}} OTF(\mathbf{r}+\boldsymbol{\Delta})|f(\mathbf{r})|^2 d\mathbf{r} \quad (4)$$
$$= \iiint_{\mathbf{r}} OTF(\mathbf{r})|f(\mathbf{r}-\boldsymbol{\Delta})|^2 d\mathbf{r}$$

$\boldsymbol{\Delta}$ is a vector in the transverse direction. We describe the detector as an array of infinitesimal pinholes. The light detected by an off-focus pinhole is

$$I_{\boldsymbol{\Delta}} = \iiint_{\mathbf{r}} |T_i(\mathbf{r})T_d(\mathbf{r}+\boldsymbol{\Delta})f(\mathbf{r})|^2 d\mathbf{r}$$
$$= \iiint_{\mathbf{r}} OTF_{\boldsymbol{\Delta}}(\mathbf{r})|f(\mathbf{r})|^2 d\mathbf{r} \quad (5)$$

where we define the off-focus OTF for displacement $\boldsymbol{\Delta}$ from focus, $OTF_{\boldsymbol{\Delta}}$.

In Fig. 1 we simulate the regions contributing to focus and off-focus pixels on the detector. Figure 1(a), gives $OTF_{\boldsymbol{\Delta}}$ with $\boldsymbol{\Delta} = 0$ for a system with a NA of 0.866 in air (so the solid angle spans 120 degrees). In Figures 1(b), (c), and (d) we estimate the off-focus OTF's for detector pixels at one, two, and three wavelengths, respectively, from focus.

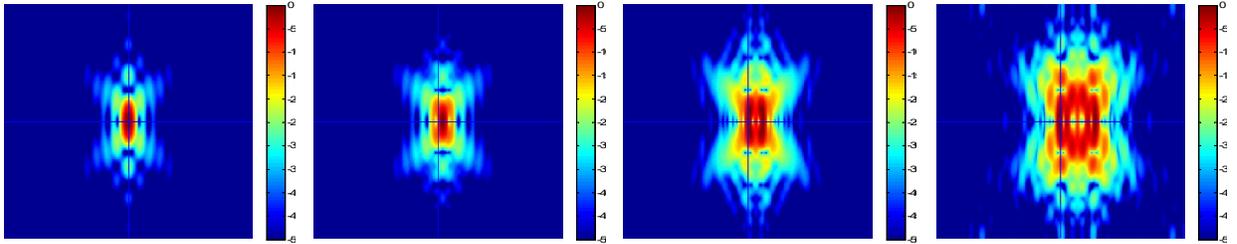

**Figure 1: x-z cross-sections of the OTF for different $\Delta$, i.e., different detector pixels. Each is normalized by its peak power. The on-focus simulation appears to agree well with the experimental result from [11].**

The shifted pixels clearly contain more information about the off-focus light, which we will use to computationally reduce the sidelobes in Figure 1(a). In Figure 2 we plot the relative amounts of total light power in the focal region (i.e. the main lobe) and outside the focus for different shifted pixels versus the shift in wavelengths.

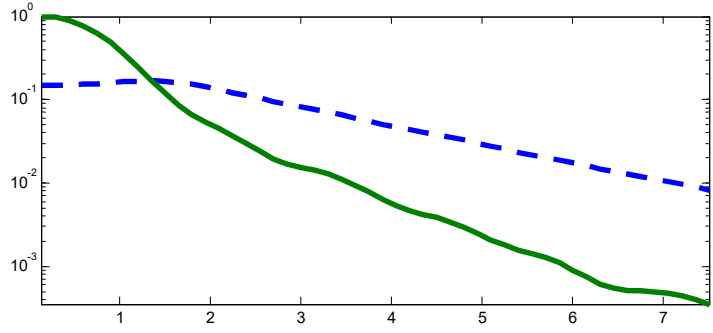

**Figure 2. Total focal power (solid) and total out-of-focus power (dashed) versus shift in wavelengths.**

Now we derive a method to improve the image using the off-focus pixels. We consider a linear combination of detector pixels

$$\begin{aligned} I_c &= \sum_n c_n I_{\Delta_n} \\ &= \sum_{k'} c_{k'} \iiint_\mathbf{r} OTF_{\Delta_n}(\mathbf{r}) |o(\mathbf{r})|^2 d\mathbf{r} \\ &= \iiint_\mathbf{r} \left\{ \sum_n c_n OTF_{\Delta_n}(\mathbf{r}) \right\} |o(\mathbf{r})|^2 d\mathbf{r} \\ &= \iiint_\mathbf{r} COTF(\mathbf{r}) |o(\mathbf{r})|^2 d\mathbf{r} \end{aligned} \quad (6)$$

where we have defined a "computational OTF", $COTF(\mathbf{r})$.

Our goal is to produce a computational OTF with the best focusing properties possible. To do this we optimize the relative amounts light it has from points in focus and out-of-focus. We discretize the problem by sampling the COTF over a grid of points, defining vectors

$$\mathbf{t}_n = \begin{pmatrix} OTF_{\Delta_n}(\mathbf{r}_1) \\ OTF_{\Delta_n}(\mathbf{r}_2) \\ \vdots \\ OTF_{\Delta_n}(\mathbf{r}_K) \end{pmatrix}. \quad (7)$$

with the different OTF's for different detector pixels from Eq. (5). The computational OTF is a linear combination of these which we write as a combination of vectors for the transfer functions for different pixels,

$$\begin{aligned} \mathbf{t}^{(c)} &= \begin{pmatrix} \mathbf{t}_0 & \mathbf{t}_1 & \cdots & \mathbf{t}_N \end{pmatrix} \mathbf{c} \\ &= \mathbf{T}\mathbf{c} \end{aligned} \quad (8)$$

Bold lower-case letters denote vectors and upper-case letters denote matrices. We will consider different choices for the pixels to use in this combination. We separate the focal/out-of-focus light with vectors **f** and **g**, where $f_i = 1$ for samples $i$ in the focal region and $f_i = 0$ for samples $i$ out-of-focus, and where $g_i = 1 - f_i$, for the same samples.

To optimize the COTF we simply maximize the ratio of focus to out-of-focus light. Since the computational OTF may be negative, we use the squared modulus of the COTF values in the optimization. Then the maximization of this ratio forms the generalized eigenvalue problem

$$\max_{\mathbf{c}} \frac{\sum_j [(\mathbf{Tc})_j f_j]^2}{\sum_j [(\mathbf{Tc})_j g_j]^2} = \max_{\mathbf{c}} \frac{\mathbf{c}^T \mathbf{A} \mathbf{c}}{\mathbf{c}^T \mathbf{B} \mathbf{c}} \tag{9}$$

The matrices **A** and **B** are

$$\begin{aligned} \mathbf{A} &= \mathbf{T}^T \mathrm{diag}(\mathbf{f}) \mathbf{T} \\ \mathbf{B} &= \mathbf{T}^T \mathrm{diag}(\mathbf{g}) \mathbf{T} \end{aligned} \tag{10}$$

where $\mathrm{diag}(\mathbf{f})$ and $\mathrm{diag}(\mathbf{g})$ are diagonal matrices with **f** and **g** on the main diagonal, respectively, and "$^T$" denotes the matrix transpose. We can regularize this result by truncating the eigenvectors of **T** to those corresponding to eigenvalues above a threshold.

## Simulation Results

First we consider the degree of improvement possible using a 3D system where a 2D detector array captures the off-focus points on a grid. We define the focal light as falling within the mainlobe of the regular OTF. Figure 3 shows examples of **c** for the 120-degree aperture system (NA 0.866 in air) with no regularization (a), and with the eigenvectors truncated at 30 dB (b) and 20 dB (c). These images are the elements of the linear combination vector arranged into a grid corresponding to the detector pixels which they weight, with the conventional pinhole location at the center pixel. The unregularized system predominantly weights weaker off-pixel detected values, potentially amplifying noise to an undesirable degree. The regularized results predominantly use the focal pinhole value and adjacent pixels, and should be much more robust.

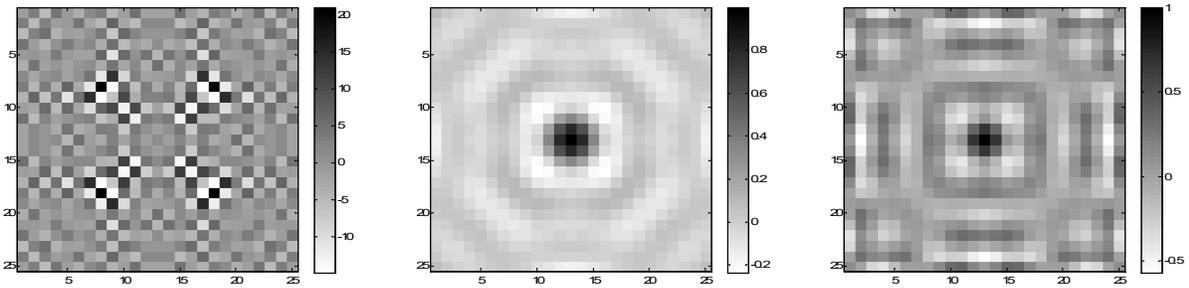

**Figure 3.** Coefficients for formation of computational estimate. Image dimensions are 3 wavelengths on a side.

The optimal ratios between focal and out-of-focus light do not vary significantly with regularization. For the unregularized case, the improvement is a factor of 7.14 over the pinhole. With truncation of eigenvectors at 30 dB it drops to 6.29, and truncation at 20 dB drops it to 5.95. So it would seem a reasonable trade-off can be made between robustness and optimality.

We construct the computational OTF using the optimal combination for the 20 dB truncation in Figure 3, and can see how the sidelobes have been reduced.

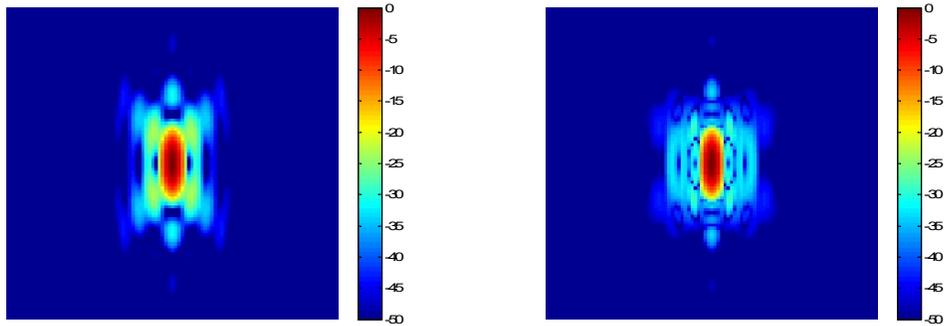

**Figure 4. original (left) versus computational (right) OTF.**

If we consider the light in planes at different depth in the OTF, shown in Figure 5, we see a modest improvement. The ratio between the two estimates is given in Figure 6.

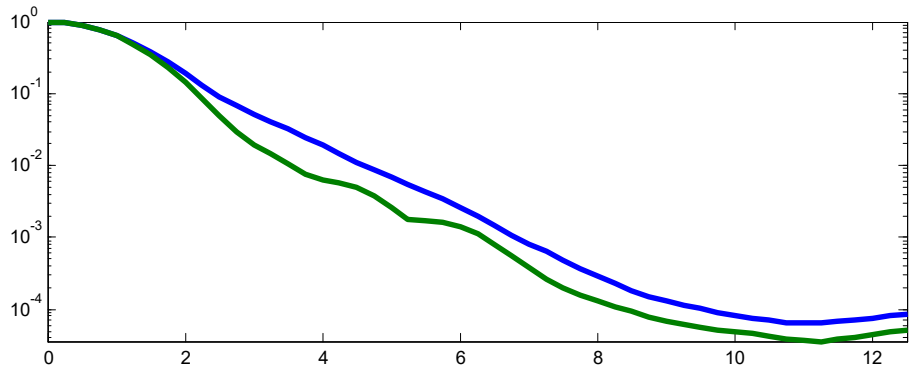

**Figure 5. Sum of light over planes of increasing defocus for the original and computational OTF.**

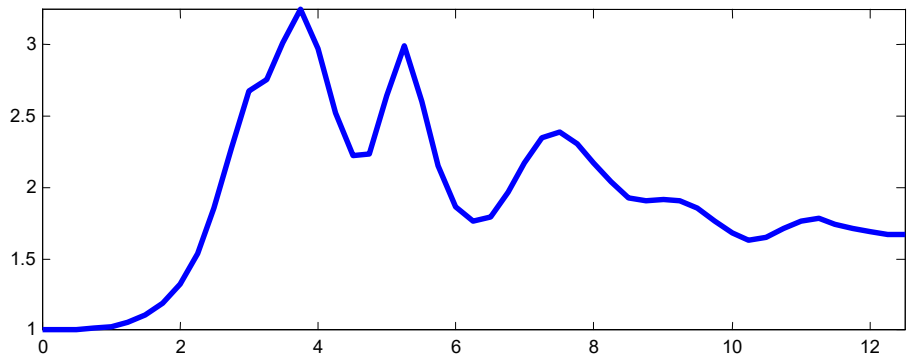

**Figure 6. Ratio of total light over planes of increasing defocus.**

Note that the sum over absolute value of the OTF must be taken for the computational OTF.

We briefly consider an example of estimating depth with this technique. Instead of maximizing light in the mainlobe, we maximize light one wavelength axial distance from the focal plane. The filter and resulting OTF are shown in Figure 7.

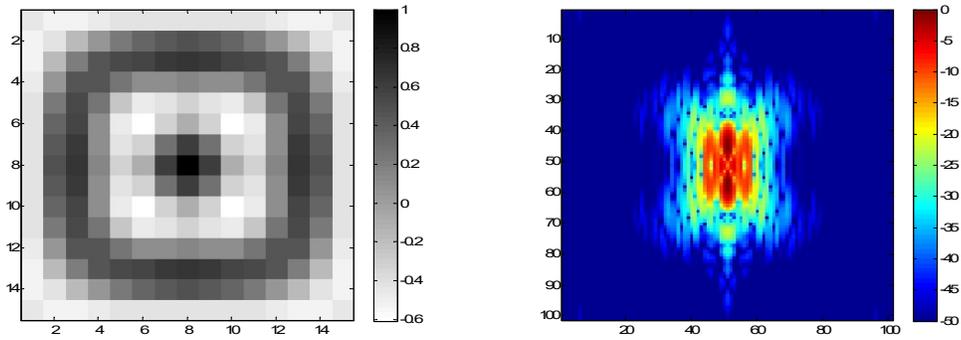

**Figure 7. Ratio of total light over planes of increasing defocus.**

The depth symmetry of the transfer functions leads to a result that is necessarily symmetric about the focal plane, but aside from that we are able to computationally produce a reasonable "focus" off the focal plane. This suggests the possibility of performing full 3D volume imaging using by computationally estimating multiple depths simultaneously during a scan over a single focal plane. Though an asymmetry in depth is needed, for example by aberrating the optical point-spread function.

In Figure 8 we show estimates over different angular aperture sizes for the unregularized, and regularized to 30 dB, 20 dB, and 10 dB.

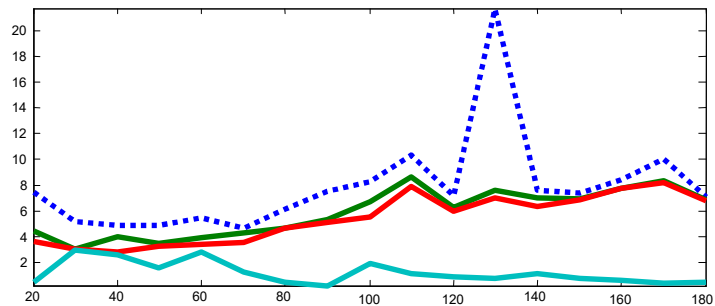

**Figure 8. Factor of improvement between computational and conventional ratios of focal to out-of-focus light.**

Aside from a spurious result with the unregularized optimal which is likely not robust, we see that performance doesn't change much with numerical aperture.

**Line Scanning**

Next we consider the potential improvement that can be achieved with a line scanning system. In this case we have more potential for improvement, and less data to contend with. Here we consider a detector with multiple line detectors, and assume each are infinitesimally narrow.

First we estimate the optimal ratio of out-of-focus light for a 0.866 NA system using the neighboring detector lines, and find that while there should be a greater potential for improvement over the point-scan system, the result is roughly the same as the point-scanning case. Coefficients for an example is shown in Figure 9, where the improvement ratio with the computational technique is 6.88, 6.51, 6.02, and 5.11 for unregularized and SVD truncated to 30 dB, 20 dB, and 10 dB respectively.

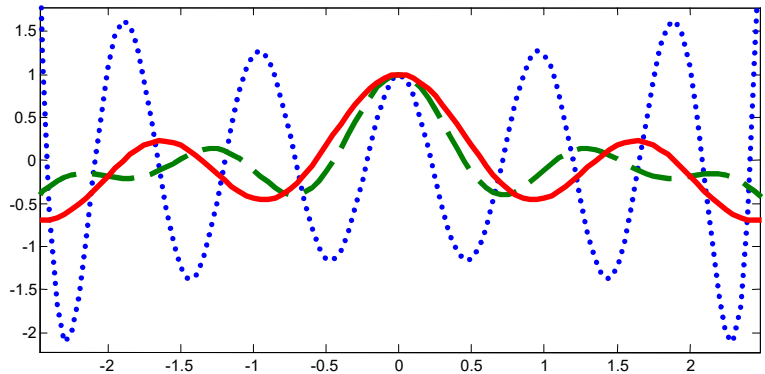

**Figure 9. Optimal coefficients for unregularized (dotted) and regularized to 30 dB (dashed), and 20 dB (solid). x-axis normalized to wavelength.**

Note that for the regularized results, the optimal combination subtracts a fraction of the off-center lines from the central line, roughly similar to the combination used in the technique demonstrated in [7].

Next we consider using the off-focus lines of neighboring steps in the scan. Recall we previously had been considering an illumination transfer function which was focused on the pinhole and a detection transfer function focused a different point on the detector. Now we include products which, relative to each pinhole location in the scan, can be treated as "cross-shifts". We combine an off-focus transfer function from the illumination and an off-focus transfer function for the detector. The total set of OTF's to be used in the optimization is two-dimensional, one dimensions for illumination shifts and the other for detector pixel locations. This amounts to the total data collected over the entire scan by stacking all the line images together.

In Figure 10, we find that using this data set to optimize the out-of-focus light ratio we can increasingly-large improvement, limited by the amount of regularization that is needed.

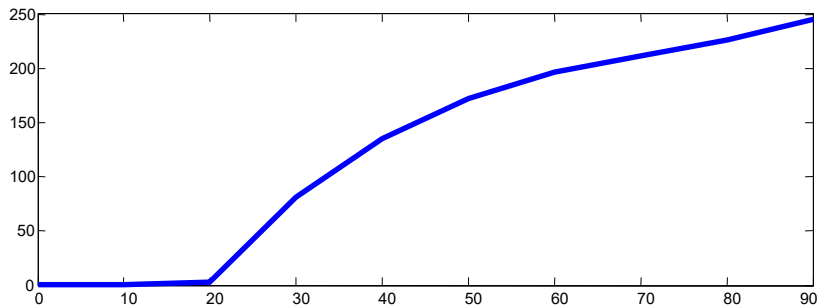

**Figure 10. Plot of improvement factor in out-of-focus light ratio compared to conventional line scan estimate for different levels of regularization in dB.**

So for this case, in a real system, care taken in determining the appropriate cutoff point would be valuable for maximizing the improvement gained. The resulting computational OTF for the 30 dB case is given in Figure 11, compared to the original OTF.

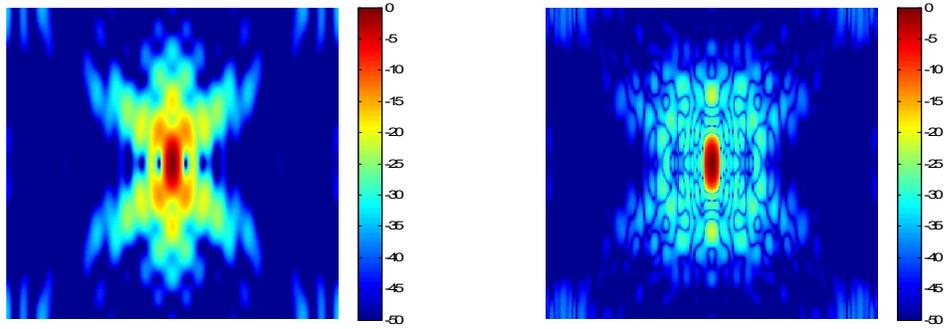

**Figure 11. Original OTF (left) versus computational OTF (right).**

We simulate the total light over planes of increasing defocus in Figure 12, and the ratio in figure 13.

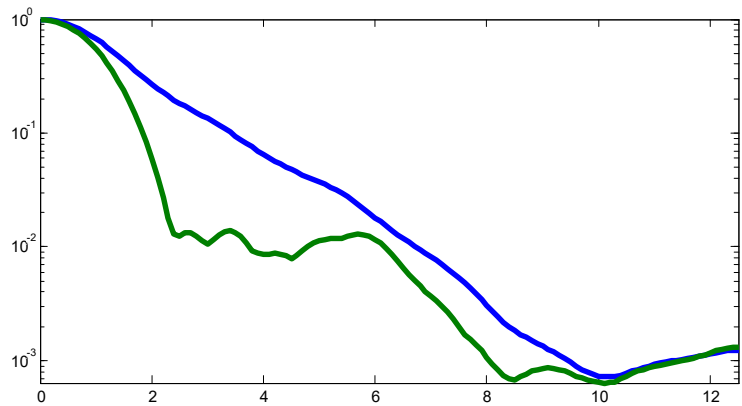

**Figure 12. Sum of light over planes of increasing defocus for the original and computational OTF.**

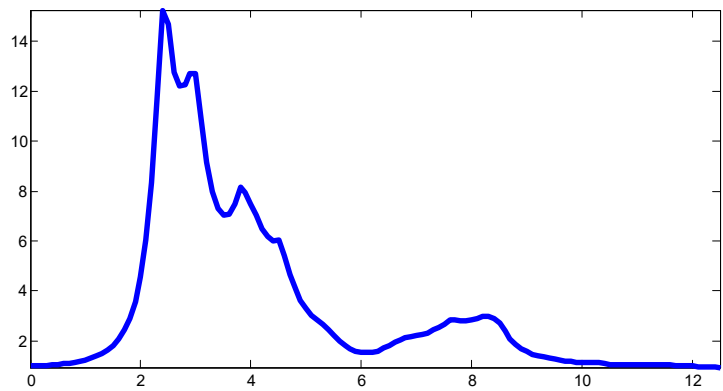

**Figure 13. Ratio of total light over planes of increasing defocus.**

## Conclusion

We estimated the amount of improvement that may be attained by using additional detectors in a confocal system. We assumed the ability to accurately model the 3D point spread function using the Debye approximation. We found that a reduction in out-of-focus light by a factor of roughly six could be achieved using a single image per step in the scan. A greater improvement may be possible by using the combined data from different steps in the scan, as was estimated for a line-scanning system. For the point-scanning system the full set of data spans four dimensions (one detection image per pixel) so the storage and computation with this set would be challenging. Still the linescan results suggests additional effort should lead to increasing performance.

These results provide an upper limit on what may be achieved, by giving the optimal improvement without much regard for detector noise or errors in the OTF estimates. Other potentially problematic issues include sample-induced aberrations and stray light, which would cause the description of the illumination to be incorrect. These problems are mitigated by regularization, as the coefficients become smooth and peaked at the center (i.e., conventional pinhole signal) providing a graceful degradation with noise and model errors

Here we basically focused on reducing the sidelobe level, but a similar approach could be taken to reducing the mainlobe width. The variation in the mainlobe component in Figure 1 suggests improvement may also be possible. We also note that this approach might be used to address the trade-off between light level and resolution in choosing pinhole size. With an array detector, we can use large pixels which would collect more light but capture blurred versions of the OTF. Then we mitigate this degradation computationally. Further the solution with blurred OTF's would be more robust against noise and model errors.

## References

1. José-Angel Conchello and Jeff W Lichtman, "Optical sectioning microscopy", *Nature Methods* 2: 920-931 (2005).
2. Keith Dillon, and Yeshaiahu Fainman, "Computational confocal tomography for simultaneous reconstruction of objects, occlusions, and aberrations," *Applied Optics* 49: 2529-2538.
3. Keith Dillon and Yeshaiahu Fainman, "Depth sectioning of attenuation", *JOSA A*, 27: 1347-1354 (2010).
4. AS Goy, M Unser, D Psaltis, "Multiple contrast metrics from the measurements of a digital confocal microscope", *Biomedical optics express* 7: 1091-1103, 2013.
5. Damien Loterie, Salma Farahi, Ioannis Papadopoulos, Alexandre Goy, Demetri Psaltis, and Christophe Moser, "Digital confocal microscopy through a multimode fiber", *Optics Express* 23: 23845-23858 (2015).
6. Wes Wallace, Lutz H. Schaefer, and Jason R. Swedlow, "A Workingperson's Guide to Deconvolution in Light Microscopy,", *BioTechniques* 31: 1076-1097 (2001).
7. Vincent Poher, Gordon T. Kennedy, Hugh B. Manning, Dylan M. Owen, Haoxiang X. Zhang, Erdan Gu, Martin D. Dawson, Paul M. W. French, and Mark A. A. Neil, "Improved sectioning in a slit scanning confocal microscope," *Optics Letters* 33: 1813-18115 (2008).
8. Changgeng Liu and Myung K. Kim, "Digital adaptive optics line-scanning confocal imaging system", *Journal of biomedical optics* 20: 111203 (2015).
9. Michael Hughes and Guang-Zhong Yang, "Line-scanning fiber bundle endomicroscopy with a virtual detector slit", Biomedical optics express 7: 2257-2268 (2016).
10. Min Gu, *Advanced Optical Imaging Theory*, Springer (2000).

11. Michael J. Nasse, Jörg C. Woehl, and Serge Huant, "High-resolution mapping of the three-dimensional point spread function in the near-focus region of a confocal microscope," *Appl. Phys. Lett.* 90: 031106 (2007).